\documentclass{article}
\usepackage[utf8]{inputenc}
\usepackage[T2A]{fontenc}
\usepackage[english]{babel}
\inputencoding{utf8}

\usepackage{amsmath, amssymb, amsthm, nicefrac, mathtools, url}

\usepackage[
    pdfpagemode=UseNone,
    bookmarks=true,
    bookmarksopen=true,
    pdfpagelayout=SinglePage,
    pdfstartview={XYZ null null 1},
    pdfhighlight=/P,
    colorlinks=true,
    linkcolor=blue,
    citecolor=blue,
    urlcolor=blue
    ]{hyperref}
\usepackage{blkarray}
\usepackage{float}

\advance\textwidth 20mm  \advance\oddsidemargin -10mm
\advance\textheight 20mm \advance\topmargin -15mm

\theoremstyle{plain}

\newtheorem{remark}{Remark}

\DeclareMathOperator{\const}{const}
\DeclareMathOperator{\diag}{diag}
\DeclareMathOperator{\Mat}{Mat}

\newcommand{\brackets}[1]{\left( #1 \right)}

\newcommand{\blue}[1]{{\color{blue} #1}}

\DeclareMathOperator{\PII}{P_{2}}
\DeclareMathOperator{\PIV}{P_{4}}
\newcommand{\Painleve}{Painlev{\'e} }
\newcommand{\PPainleve}{Painlev{\'e}}
\newcommand{\PIVn}[1]{\text{P}_4^{ #1 }}

\mathtoolsset{showonlyrefs,showmanualtags}

\title{On matrix Painlev\'e-4 equations. \\
Part 2: Isomonodromic Lax pairs}

\date{13 August 2021}

\author{I.A. Bobrova\thanks{National Research Univerisity ``Higher School of Economics'', Moscow, Russian Federation.},~ V.V. Sokolov\thanks{L.D.~Landau Institute for Theoretical Physics, Chernogolovka, Russian Federation.} \thanks{Federal University of ABC, Santo Andr\'e, Sao Paulo, Brazil. E-mail: vsokolov@landau.ac.ru}}

\usepackage{tikz}

\begin{document}
\maketitle

\begin{abstract}
For all non-equivalent matrix systems of Painlev\'e-4 type found by authors in \cite{Bobrova_Sokolov_2020_1},  isomonodromic Lax pairs are presented. Limiting transitions from these systems to matrix  Painlev\'e-2 equations are found.
\medskip

\end{abstract}


\section{Introduction}

It is known that the scalar system
\begin{equation} \label{syscal}
\left\{
\begin{array}{lcl}
u' &=& -u^2 + 2 u v - 2 z u + c_1 , \\[2mm]
v' &=& -v^2 + 2 u v + 2 z v + c_2,
\end{array}
\right.
\end{equation}
is equivalent to the Painlev\'e-4 equation
\begin{equation}\label{P4scal}
 y''= \frac{y'^2}{2\, y}  +\frac{3}{2} y^3+4z y^2 +2(z^2-\gamma) y+\frac{\delta}{y},
\end{equation}
where \,\, $y(z) = u(z)$, \,\, $\gamma=1 + \frac{1}{2} c_1 - c_2$, \,\, $\delta = - \frac{1}{2} c_1^2$. The prime $'$ means the derivative with respect to  the variable $z \in \mathbb{C}$.

In the paper \cite{Bobrova_Sokolov_2020_1}, matrix generalizations of system \eqref{syscal} of the form 
\begin{equation}\label{3rootTail}
\left\{
\begin{array}{lcl}
u' &=&  - u^2 + 2 \,u v  + \alpha (u v - v u) - 2 z u + b_1 u+u b_2+b_3 v+v b_4+b_5\\ [2mm]
v' &=& -  v^2 + 2 \, v u + \beta (v u - u v) + 2 z v+c_1 v+v c_2 +c_3 u +u c_4+c_5,
\end{array}   
\right.
\end{equation}
were investigated. In system \eqref{3rootTail} the coefficients $\alpha$, $\beta$ are scalar and others are constant $n \times n$-matrices. 
Here and below, we consider matrices over the field $\mathbb{C}$. The authors found all non-equivalent systems of the form \eqref{3rootTail} that satisfy the matrix \PPainleve--Kovalevskaya test~\cite{Balandin_Sokolov_1998}.
They can be written as follows:
\begin{align} \label{eq:case1tail_can}
    \tag*{$\PIVn{0}$}
    &\left\{
    \begin{array}{lcl}
         u' 
         &=& - u^2 + u v + v u - 2 z u 
         + h_1 u 
         + \gamma_1 \, \mathbb{I},
         \\[2mm]
         v' 
         &=& - v^2 + v u + u v + 2 z v
         - v h_1
         + \gamma_2 \, \mathbb{I},
    \end{array}
    \right.
    \hspace{1.15cm}
\end{align}
\begin{align} 
    \label{eq:case3tail_can}
    \tag*{$\PIVn{1}$}
    &\left\{
    \begin{array}{lcl}
         u' 
         &=& - u^2 + 2 u v - 2 z u 
         + h_2,
         \\[2mm]
         v' 
         &=& - v^2 + 2 u v + 2 z v
         + h_2 + \gamma_1 \, \mathbb{I},
    \end{array}
    \right.
    \hspace{2.cm}
\end{align}
\begin{align} 
    \label{eq:case5tail_can}
    \tag*{$\PIVn{2}$}
    &\left\{
    \begin{array}{lcl}
         u' 
         &=& - u^2 + 2 u v - 2 z u 
         + h_2,
         \\[2mm]
         v' 
         &=& - v^2 + 3 u v - v u + 2 z v
         + h_1 u 
         + 2 h_2 + \gamma_1 \, \mathbb{I}.
    \end{array}
    \right.
\end{align}
Here $\gamma_1$, $\gamma_2 \in \mathbb{C}$, $h_i$ are arbitrary matrices and two constant matrices $h_1$, $h_2$ in \ref{eq:case5tail_can} are connected by the relation $[h_2, h_1] = - 2 h_1$. 

In the present paper we construct for each of systems \ref{eq:case1tail_can} -- \ref{eq:case5tail_can} an isomonodromic Lax representation of the form
\begin{equation}\label{eq:isomLax}
    A_z=B_\zeta+[B,A],
\end{equation}
where $\zeta$ is the spectral parameter, $A(z, \zeta)$,\, $B(z, \zeta)$ are some $2\times2$-matrices with entries being polynomials in $u$, $v$, $z$. We assume that the $\zeta$-dependence of $A$ and $B$ is given by
\begin{align}\label{AABB}
    A
    &=A_1 \zeta +A_0 +A_{-1} \zeta^{-1}, 
    &
    B
    &= B_1 \zeta + B_0,
\end{align}
where $A_1$ and $B_1$ are constant diagonal matrices.

\begin{remark}\label{Rem1}
It is clear that in the scalar case a pair of the form \eqref{AABB} admits the shift
\begin{align}\label{shift}
    A
    &\mapsto A+\kappa_1 \zeta+\kappa_2+\kappa_3 \zeta^{-1}+\kappa_4 z, 
    &
    B
    &\mapsto B+\kappa_4 \zeta +\kappa_5 f(u,v,z) + \kappa_6,
\end{align}
where $\kappa_i$ are arbitrary constant scalar matrices, $f(u,v,z)$ is an arbitrary function. In addition, the scaling $\zeta \mapsto \const \, \zeta$ and a conjugation of $A$ and $B$ by a constant diagonal matrix are allowed.
\end{remark}

Section \ref{limit} is devoted to limiting transitions from the matrix \PPainleve-4 systems \ref{eq:case1tail_can} -- \ref{eq:case5tail_can}  to the matrix \PPainleve-2 equations \ref{P20} -- \ref{P22} found in \cite{Adler_Sokolov_2020_1}. Extending these transitions to the Lax pairs from Section \ref{sect4}, we obtain Lax pairs for the \PPainleve-2 equations.

\section{Isomonodromic Lax pairs for systems \texorpdfstring{\ref{eq:case1tail_can} -- \ref{eq:case5tail_can}}{P40 -- P42}}\label{sect4}

The $\text{P}_4$ equation \eqref{P4scal} 
admits several equivalent isomonodromic representations \eqref{eq:isomLax} with different dependence on the spectral parameter $\zeta$ (see for example \cite{joshi2007linearization}). 

An isomonodromic pair of the form \eqref{AABB} for the scalar equation \eqref{P4scal} was found in \cite{Jimbo_Miwa_1981}. More precisely, the matrices 
\begin{align*}
    B (z, \zeta)
    &= 
    \begin{pmatrix}
    1 & 0 
    \\[0.9mm]
    0 & - 1
    \end{pmatrix}
    \zeta 
    + 
    \begin{pmatrix}
    0 & w 
    \\[0.9mm]
    2 (u - \theta_1) w^{-1} & 0
    \end{pmatrix}
    ,
    \\[2mm]
    A (z, \zeta)
    &=
    \begin{pmatrix}
    1 & 0 
    \\[0.9mm]
    0 & - 1
    \end{pmatrix}
    \zeta 
    + 
    \begin{pmatrix}
    z & w 
    \\[0.9mm]
    2 (u - \theta_1) w^{-1} & - z
    \end{pmatrix}
    +
    \begin{pmatrix}
    - u + \theta_0 & - \frac12 v w 
    \\[0.9mm]
    2 u (u - 2 \theta_0) v^{-1} w^{-1} & z - \theta_0
    \end{pmatrix}
    \zeta^{-1}
\end{align*}
define the representation \eqref{AABB} for the system
\begin{gather*}
    \begin{cases}
    u' 
    &= - 2 u^2 v^{-1} - u v + 4 \theta_0 u v^{-1} + \theta_1 v,
    \\[2mm]
    v'
    &= - 4 u + v^2 + 2 z v + 4 \theta_0,
    \\[2mm]
    w'
    &= - w (v + 2 z),
    \end{cases}
\end{gather*}
which is equivalent to the $\PIV$ equation \eqref{P4scal} for $y (z) = v (z)$. The key point of our approach is the observation that the transformation
\begin{align*}
    u 
    &\mapsto 2 u v,
    &
    z
    &\mapsto - z,
    &
    B 
    &\mapsto g \, B^T \, g^{-1} + g' g^{-1},
    &
    A
    &\mapsto - g \, A^T \, g^{-1},
\end{align*}
where $g = \diag\left( w^{- \frac12}, w^{\frac12} \right)$, brings this Lax pair to a polynomial pair for system \eqref{syscal} with 
\begin{gather} \label{eq:scalP4_Laxpair}
\begin{aligned}
    B (z, \zeta)
    &= 
    \begin{pmatrix}
    1 & 0 
    \\[0.9mm] 
    0 & -1
    \end{pmatrix}
    \zeta 
    + 
    \begin{pmatrix}
    \frac12 v - z & u v + c_1 
    \\[0.9mm]
    1 & - \frac12 v + z
    \end{pmatrix}
    ,
    \\[2mm]
    A (z, \zeta)
    &= 
    \begin{pmatrix}
    -1 & 0 
    \\[0.9mm] 
    0 & 1
    \end{pmatrix}
    \zeta 
    + 
    \begin{pmatrix}
    z & - u v - c_1 
    \\[0.9mm]
    -1 & -z
    \end{pmatrix}
    + \frac12 
    \begin{pmatrix}
    u v + \frac12 c_2 & - u^2 v - c_2 u 
    \\[0.9mm]
    v & - u v - \frac12 c_2
    \end{pmatrix}
    \zeta^{-1}
    .
\end{aligned}
\end{gather}

As a result of the shift $z \mapsto z + \tau$ we obtain a pair for the system 
\begin{gather}\label{sysshift}
    \begin{cases}
    u'
    &= - u^2 
    + 2 u v - 2 z u
    + 2 \tau u
    + c_1
    ,
    \\[2mm]
    v'
    &= - v^2 + 2 u v + 2 z v
    - 2 \tau v
    + c_2
    .
    \end{cases}
\end{gather}
with matrices $A$ and $B$ given by 
\begin{align} 
    B (z, \zeta)
    &= 
    \begin{pmatrix}
    1 & 0 
    \\[0.9mm] 
    0 & -1
    \end{pmatrix}
    \zeta 
    + 
    \begin{pmatrix}
    r_0 v 
    + r_1 u
    - z 
    + \tau +\kappa_6
    & 
    u v 
    + c_1 
    \\[0.9mm]
    1 
    & 
    (r_0 - 1) v 
    + r_1 u
    + z - \tau+\kappa_6
    \end{pmatrix}
    ,
    \\[-4mm]
    \label{eq:JMpairscal}
    \\[1mm]
    A (z, \zeta)
    &= 
    \begin{pmatrix}
    -1 & 0 
    \\[0.9mm] 
    0 & 1
    \end{pmatrix}
    \zeta 
    + 
    \begin{pmatrix}
    z - \tau+\kappa_2
    & 
    - u v 
    - c_1 
    \\[0.9mm]
    -1 & - z + \tau+\kappa_2
    \end{pmatrix}
    + \frac12 
    \begin{pmatrix}
    u v 
    + \frac12 c_2 
    + \kappa_3
    & 
    - u^2 v 
    - c_2 u 
    \\[0.9mm]
    v
    & 
    - u v 
    - \frac12 c_2
    + \kappa_3
    \end{pmatrix}
    \zeta^{-1}
    .
\end{align}
The parameters $r_0,r_1,\kappa_2, \kappa_3, \kappa_6$ were implemented by a shift described in Remark \ref{Rem1}. 

We use a procedure of non-abelinization of the latter pair to the matrix case. Namely, we form an ansatz for the matrix pair replacing parameters and variables $u$ and $v$ by non-commutative ones such that in the case of $1 \times 1$-matrices this ansatz coincides with \eqref{eq:JMpairscal}. In particular, we replace the commutative product $u v$ by $r \, u v \, + \, (1 - r) \, v u$ with unknown number $r$ and so on.

Note that when the parameters in \eqref{eq:JMpairscal} become matrices, no shifts like \eqref{shift} can remove them.

\subsection{Case \texorpdfstring{\,\,\ref{eq:case1tail_can}}{alpha = beta = -1}} 

In the case of $1\times 1$-matrices the \ref{eq:case1tail_can} system has the form \eqref{sysshift}, where $\tau = \frac12 h_1$, $c_1 = \gamma_1$, $c_2 = \gamma_2$. Thus, the non-abelian ansatz becomes
\begin{gather}
\begin{aligned}
    B (z, \zeta)
    &= 
    \begin{pmatrix}
    \mathbb{I} & 0 
    \\[0.9mm] 
    0 & - \mathbb{I}
    \end{pmatrix}
    \zeta 
    + 
    \begin{pmatrix}
    r_0 v 
    + r_1 u
    - z \, \mathbb{I}
    + \frac{1}{2} h_1 + \kappa_6 
    & 
    r_2 u v + (1 - r_2) v u
    + \gamma_1 \, \mathbb{I}
    \\[0.9mm]
    \mathbb{I} 
    & 
    (r_0 - 1) v 
    + r_1 u
    + z \, \mathbb{I}
    - \frac{1}{2} h_1 + \kappa_6
    \end{pmatrix}
    ,
    \\[2mm]
    A (z, \zeta)
    &= 
    \begin{pmatrix}
    - \mathbb{I} & 0 
    \\[0.9mm] 
    0 & \mathbb{I}
    \end{pmatrix}
    \zeta 
    + 
    \begin{pmatrix}
    z \, \mathbb{I}
    - \frac{1}{2} h_1 + \kappa_2
    & 
    - r_3 u v - (1 - r_3) v u
    - \gamma_1 \, \mathbb{I}
    \\[0.9mm]
    - \mathbb{I} & - z \, \mathbb{I}
    + \frac{1}{2} h_1 + \kappa_2
    \end{pmatrix}
    \\[1mm] 
    & \qquad
    + \frac12 
    \begin{pmatrix}
    r_4 u v + (1 - r_4) v u
    + \frac12 \gamma_2 \, \mathbb{I}
    & 
    - r_6 u^2 v 
    - r_7 u v u
    - (1 - r_6 - r_7) v u^2
    - \gamma_2 u 
    \\[0.9mm]
    v
    & 
    - r_5 u v - (1 - r_5) v u
    - \frac12 \gamma_2 \, \mathbb{I}
    \end{pmatrix}
    \zeta^{-1}
    .
\end{aligned}
\end{gather}

Substituting it in \eqref{AABB}, we obtain two possible Lax pairs given by 
\begin{align*}
    &&  \kappa_2
    &= \phantom{-} \frac{h_1}{2}, 
    &
    \kappa_6
    &=\frac{h_1}{2}, 
    &
    r_0
    &= r_2 = r_3 = r_4 = r_7
    = 1,
    &
    r_1
    &= r_5 = r_6
    = 0,
    \\
    &\text{or}
    & \kappa_2
    &=-\frac{h_1}{2}, 
    &
    \kappa_6
    &=\frac{h_1}{2}, 
    &
    r_4 
    &= r_7
    = 1,
    &
    r_0 
    &= r_1 = r_2 = r_3 = r_5 = r_6
    = 0.
 \end{align*}
In the first case the matrices $B$ and $A$ read
\begin{gather} \label{eq:case1_Laxpair}
\begin{aligned}
    B (z, \zeta)
    &= 
    \begin{pmatrix}
    \mathbb{I} & 0 
    \\[0.9mm] 
    0 & - \mathbb{I}
    \end{pmatrix}
    \zeta 
    + 
    \begin{pmatrix}
    v 
    - z \, \mathbb{I}
    + h_1
    & 
    u v
    + \gamma_1 \, \mathbb{I}
    \\[0.9mm]
    \mathbb{I} 
    & 
    z \, \mathbb{I}
    \end{pmatrix}
    ,
    \\[2mm]
    A (z, \zeta)
    &= 
    \begin{pmatrix}
    - \mathbb{I} & 0 
    \\[0.9mm] 
    0 & \mathbb{I}
    \end{pmatrix}
    \zeta 
    + 
    \begin{pmatrix}
    z \, \mathbb{I}
    & 
    - u v
    - \gamma_1 \, \mathbb{I}
    \\[0.9mm]
    -\mathbb{I} & - z \, \mathbb{I} + h_1
    \end{pmatrix}
    + \frac12 
    \begin{pmatrix}
    u v
    + \frac12 \gamma_2 \, \mathbb{I}
    & 
    - u v u
    - \gamma_2 u 
    \\[0.9mm]
    v
    & 
    - v u
    - \frac12 \gamma_2 \, \mathbb{I}
    \end{pmatrix}
    \zeta^{-1}
    .
\end{aligned}
\end{gather}
To get the second pair, one can apply the following symmetry of the \ref{eq:case1tail_can} system:
\begin{align}
    u
    &\mapsto e^{- z h_1} u^{T} e^{z h_1},
    &
    v
    &\mapsto e^{- z h_1} v^{T} e^{z h_1}
\end{align}
to \eqref{eq:case1_Laxpair} and then conjugate the resulting Lax pair by the matrix $s~=~\small \begin{pmatrix} 0 & - \mathbb{I} \\ \mathbb{I} & \phantom{-} 0 \end{pmatrix}$ to bring it to the form \eqref{eq:JMpairscal}.

\medskip

\subsection{Case \texorpdfstring{\,\,\ref{eq:case3tail_can}}{alpha =0, beta = -2}} 

In the case $\alpha = 0$, $\beta = -2$  we have $\tau = 0$, $c_1 = h_2$, $c_2 = h_2 + \gamma_1$ and the ansatz is given by 
\begin{gather}
\begin{aligned}
    B (z, \zeta)
    &= 
    \begin{pmatrix}
    \mathbb{I} & 0 
    \\[0.9mm] 
    0 & -\mathbb{I}
    \end{pmatrix}
    \zeta 
    + 
    \begin{pmatrix}
    r_0 v 
    + r_1 u
    - z \, \mathbb{I}
    & 
    r_2 u v + (1 - r_2) v u
    + h_2
    \\[0.9mm]
    \mathbb{I} 
    & 
    (r_0 - 1) v 
    + r_1 u
    + z \, \mathbb{I}
    \end{pmatrix}
    ,
    \\[2mm]
    A (z, \zeta)
    &= 
    \begin{pmatrix}
    -\mathbb{I} & 0 
    \\[0.9mm] 
    0 & \mathbb{I}
    \end{pmatrix}
    \zeta 
    + 
    \begin{pmatrix}
    z \, \mathbb{I}
    & 
    - r_3 u v - (1 - r_3) v u
    - h_2
    \\[0.9mm]
    -\mathbb{I} & - z \, \mathbb{I}
    \end{pmatrix}
    \\[1mm]
    & \qquad
    + \frac12 
    \begin{pmatrix}
    \begin{array}{c}
    r_4 u v + (1 - r_4) v u
    \\[0.5mm]
    + \frac12 h_2
    + \frac12 \gamma_1 \, \mathbb{I}
    + \kappa_3
    \end{array}
    & 
    \begin{array}{c}
    - r_6 u^2 v 
    - r_7 u v u
    - (1 - r_6 - r_7) v u^2
    \\[0.5mm]
    - r_8 h_2 u - (1 - r_8) u h_2
    - \gamma_1 u 
    \end{array}
    \\[4.2mm]
    v
    & 
    - r_5 u v - (1 - r_5) v u
    - \frac12 h_2
    - \frac12 \gamma_1 \, \mathbb{I}
    + \kappa_3
    \end{pmatrix}
    \zeta^{-1}
    .
\end{aligned}
\end{gather}

From \eqref{AABB} it follows that
\begin{align*}
    &&
    \kappa_3
    &= \phantom{-} \frac{h_2}{2},
    &
    r_1 
    &= r_2 = r_3 = r_4 = r_7 = r_8
    = 1,
    &
    r_0 
    &= r_5 = r_6
    = 0,
    \\
    &\text{or}
    &
    \kappa_3
    &= - \frac{h_2}{2},
    &
    r_2
    &= r_3 = r_4 = r_5 = r_6
    = 1,
    &
    r_0
    &= r_1 = r_7 = r_8
    = 0.
\end{align*}
In the second case the matrices $B$ and $A$ have the form
\begin{gather} \label{eq:case3_Laxpair}
\begin{aligned}
    B (z, \zeta)
    &=
    \begin{pmatrix}
    \mathbb{I} & 0 
    \\[0.9mm] 
    0 & -\mathbb{I}
    \end{pmatrix}
    \zeta 
    + 
    \begin{pmatrix}
    - z \, \mathbb{I}
    & 
    u v
    + h_2
    \\[0.9mm]
    \mathbb{I} 
    & 
    - v + z \, \mathbb{I}
    \end{pmatrix}
    ,
    \\[2mm]
    A (z, \zeta)
    &= 
    \begin{pmatrix}
    -\mathbb{I} & 0 
    \\[0.9mm] 
    0 & \mathbb{I}
    \end{pmatrix}
    \zeta 
    + 
    \begin{pmatrix}
    z \, \mathbb{I}
    & 
    - u v
    - h_2
    \\[0.9mm]
    -\mathbb{I} & - z \, \mathbb{I}
    \end{pmatrix}
    + \frac12 
    \begin{pmatrix}
    u v
    + \frac12 \gamma_1 \, \mathbb{I}
    & 
    - u^2 v
    - u (h_2 + \gamma_1 \, \mathbb{I})
    \\[0.9mm]
    v
    & 
    - u v
    - h_2
    - \frac12 \gamma_1 \, \mathbb{I}
    \end{pmatrix}
    \zeta^{-1}
    .
\end{aligned}
\end{gather}
The first pair should be equivalent to this, but we have not yet been able to find an appropriate gauge transformation.

\medskip
\subsection{Case \texorpdfstring{\,\,\ref{eq:case5tail_can}}{alpha =0, beta = -3}} 

The case $\alpha = 0$, $\beta = -3$ is different from the previous two since for $1\times 1$-matrices the commutation relation  $[h_2, h_1] = - 2 h_1$ implies  $h_1 = 0.$  If $h_1 = 0$ then  in the scalar equation \eqref{sysshift} we have $\tau = 0$, $c_1 = h_2$, $c_2 = 2 h_2 + \gamma_1$. Substituting the ansatz
\begin{gather}
\begin{aligned}
    B (z, \zeta)
    &= 
    \begin{pmatrix}
    \mathbb{I} & 0 
    \\[0.9mm] 
    0 & -\mathbb{I}
    \end{pmatrix}
    \zeta 
    + 
    \begin{pmatrix}
    r_0 v 
    + r_1 u
    - z \, \mathbb{I}
    & 
    r_2 u v + (1 - r_2) v u
    + h_2
    \\[0.9mm]
    \mathbb{I} 
    & 
    (r_0 - 1) v 
    + r_1 u
    + z \, \mathbb{I}
    \end{pmatrix}
    ,
    \\[2mm]
    A (z, \zeta)
    &= 
    \begin{pmatrix}
    -\mathbb{I} & 0 
    \\[0.9mm] 
    0 & \mathbb{I}
    \end{pmatrix}
    \zeta 
    + 
    \begin{pmatrix}
    z \, \mathbb{I}
    & 
    - r_3 u v - (1 - r_3) v u
    - h_2
    \\[0.9mm]
    -\mathbb{I} & - z \, \mathbb{I}
    \end{pmatrix}
    \\[1mm]
    & \qquad
    + \frac12 
    \begin{pmatrix}
    \begin{array}{c}
    r_4 u v + (1 - r_4) v u
    \\[0.5mm]
    + h_2
    + \frac12 \gamma_1 \, \mathbb{I}
    + \kappa_3
    \end{array}
    & 
    \begin{array}{c}
    - r_6 u^2 v 
    - r_7 u v u
    - (1 - r_6 - r_7) v u^2
    \\[0.5mm]
    - r_8 h_2 u - (2 - r_8) u h_2
    - \gamma_1 u 
    \end{array}
    \\[4.2mm]
    v
    & 
    - r_5 u v - (1 - r_5) v u
    - h_2 
    - \frac12 \gamma_1 \, \mathbb{I}
     + \kappa_3
    \end{pmatrix}
    \zeta^{-1}
\end{aligned}
\end{gather}
into \eqref{AABB}, we obtain
\begin{align*}
    \kappa_3
    &= 0,
    &
    r_1
    &= r_2 = r_3 = r_4 = r_5 = r_6 = r_8
    = 1,
    &
    r_0
    &= r_7
    = 0,
\end{align*}
and therefore
\begin{align}
    B (z, \zeta)
    &= 
    \begin{pmatrix}
    \mathbb{I} & 0 
    \\[0.9mm] 0 & -\mathbb{I}
    \end{pmatrix}
    \zeta 
    + 
    \begin{pmatrix}
    u - z \, \mathbb{I}
    & 
    u v
    + h_2
    \\[0.9mm]
    \mathbb{I} 
    & 
    u - v + z \, \mathbb{I}
    \end{pmatrix}
    ,
    \\[-4mm]
    \label{AB03}
    \\[1mm]
    A (z, \zeta)
    &= 
    \begin{pmatrix}
    -\mathbb{I} & 0 
    \\[0.9mm] 0 & \mathbb{I}
    \end{pmatrix}
    \zeta 
    + 
    \begin{pmatrix}
    z \, \mathbb{I}
    & 
    - u v
    - h_2
    \\[0.9mm]
    -\mathbb{I} & - z \, \mathbb{I}
    \end{pmatrix}
    + \frac12 
    \begin{pmatrix}
    u v
    + h_2
    + \frac12 \gamma_1 \, \mathbb{I}
    & 
    - u^2 v
    - h_2 u 
    - u h_2 
    - \gamma_1 u
    \\[0.9mm]
    v
    & 
    - u v
    - h_2
    - \frac12 \gamma_1 \, \mathbb{I}
    \end{pmatrix}
    \zeta^{-1}
    .
\end{align}
 
To implement $h_1$ into these matrices, we use the method of undetermined coefficients, combining it with some considerations of homogeneity.
It is clear that the right hand sides of systems \ref{eq:case1tail_can} -- \ref{eq:case5tail_can} are homogeneous polynomials with weights 
\begin{align}
    &&
    w (u)
    &= w (v)
    = w (z)
    = w (h_1)
    = 1,
    &
    w (h_2)
    &= w (\gamma_1)
    = 2.
    &&
\end{align}
In the Lax pairs found above we have 
\begin{gather}
    w \brackets{A_j \{1,1\}}
    = w \brackets{A_j \{2,2\}}
    = w \brackets{B_j \{1,1\}}
    = w \brackets{B_j \{2,2\}}
    = 1 - j,
    \\[2mm]
    \begin{aligned}
    w \brackets{A_j \{1,2\}}
    &= w \brackets{B_j \{1,2\}}
    = 2 - j, 
    &&&&&
    w \brackets{A_j \{2,1\}}
    &= w \brackets{B_j \{2,1\}}
    = - j,
    \end{aligned}
\end{gather}
where $j=0,-1$. Suppose that this is true also for the Lax pair in the case $h_1 \ne 0$. All monomials of weight $1$, $2$ and $3$ that contain $h_1$ are the following:
\begin{align*}
    1
    &: 
    &
    &h_1
    ;
    \\[1mm]
    2 
    &: 
    &
    &h_1^2,
    \\
    &&
    & z h_1, \,\,
    u h_1, \,\,
    h_1 u, \,\,
    v h_1, \,\,
    h_1 v;
    \\[1mm]
    3 
    &: 
    &
    & h_1^3,
    \\
    &&
    & z h_1^2, \,\,
    u h_1^2, \,\,
    h_1 u h_1, \,\,
    h_1^2 u, \,\,
    v h_1^2, \,\,
    h_1 v h_1, \,\,
    h_1^2 v, \\
    &&
    & z^2 h_1, \,\,
    h_1 u^2, \,\,
    u h_1 u, \,\,
    u^2 h_1, \,\,
    h_1 v^2, \,\,
    v h_1 v, \,\,
    v^2 h_1, 
    \\
    &&
    & \gamma_1 \, h_1, \,\,
    h_1 h_2, \,\,
    h_2 h_1, \,\,
    u v h_1, \,\,
    v u h_1, \,\,
    h_1 u v, \,\,
    h_1 v u, \,\,
    u h_1 v, \,\,
    v h_1 u,
    \\
    &&
    &
    {z h_1 u}, \,\,
    {z u h_1}, \,\,
    {z h_1 v}, \,\,
    {z v h_1}
    .
\end{align*}
Adding to the matrices \eqref{AB03} arbitrary linear combinations of these monomials of proper weights, we constitute a new ansatz with 57 unknown coefficients. Substituting it to \eqref{eq:isomLax} and solving the corresponding large but simple algebraic system, we obtain the Lax pair for \ref{eq:case5tail_can} with
\begin{gather} \label{eq:case5_Laxpair}
\begin{aligned}
    B (z, \zeta)
    &= 
    \begin{pmatrix}
    \mathbb{I} & 0 
    \\[0.9mm] 
    0 & -\mathbb{I}
    \end{pmatrix}
    \zeta 
    + 
    \begin{pmatrix}
    u - z \, \mathbb{I}
    & 
    u v + \blue{\frac12 h_1 u} + h_2 
    \\[0.9mm]
    \mathbb{I} 
    & 
    u - v + z \, \mathbb{I}
    \end{pmatrix},
    \\[2mm]
    A (z, \zeta)
    &= 
    \begin{pmatrix}
    - \mathbb{I} & 0 
    \\[0.9mm] 
    0 & \mathbb{I}
    \end{pmatrix}
    \zeta
    + 
    \begin{pmatrix}
    z \, \mathbb{I}
    & 
    - u v - \blue{\frac12 h_1 u} - h_2 
    \\[0.9mm]
    - \mathbb{I} & - z \, \mathbb{I}
    \end{pmatrix}
    \\[1mm]
    & \qquad
    + \frac12 
    \begin{pmatrix}
    \begin{array}{c}
         u v 
         + \blue{\frac12 h_1 u }
         + h_2 
         + \frac12 \gamma_1 \, \mathbb{I}
         \\[0.5mm]
         \phantom{\frac12}
    \end{array}
    & 
    \begin{array}{c}
         - u^2 v 
         - \blue{\frac12 u h_1 u }
         - \blue{\frac12 h_1 u v }
         + \blue{z h_1 u }
         \\[0.5mm]
         - h_2 u 
         - u h_2 
         - \gamma_1 u
         - \blue{\frac12 h_1 h_2} 
    \end{array}
    \\[4.2mm]
    v 
    & 
    - u v - \blue{\frac12 h_1 u} - h_2 - \frac12 \gamma_1 \, \mathbb{I}
    \end{pmatrix}
    \zeta^{-1}
    .
\end{aligned}
\end{gather}
Additional monomials with $h_1$ are marked in blue.

\medskip

\section{Limiting transitions in Lax pairs}
\label{limit}

In the scalar case, the scheme of degenerations of the \Painleve equations \cite{gambier1910equations} and their isomonodromic Lax pairs is well-known. In particular, there exists a limiting transition from the equation $\PIV$ to the equation $\PII$. We construct similar transitions from the systems \ref{eq:case1tail_can} -- \ref{eq:case5tail_can} and their Lax pairs found in Section \ref{sect4} to the matrix  $\PII$ equations 
\begin{alignat}{2}
    \label{P20}
    \tag*{$\text{P}_2^0$}
    & y''= 2 y^3 + x y + b y + y b + \alpha \, \mathbb{I}, 
    &\qquad
    & \alpha \in {\mathbb C},~~
    b \in \Mat_n (\mathbb{C}),
    \\[1mm]
    \label{P21}
    \tag*{$\text{P}_2^1$}
    & y''= [y,y'] + 2 y^3 + x y + a, 
    &
    & a \in \Mat_n (\mathbb{C}),
    \\[1mm]
    \label{P22}
    \tag*{$\text{P}_2^2$}
    & y''= 2 [y,y'] + 2 y^3 + x y + b y + y b + a, 
    &
    & a, b \in \Mat_n (\mathbb{C}),~~ 
    [b,a] = 2 b,
\end{alignat}
discovered in the paper \cite{Adler_Sokolov_2020_1}. As a result, we obtain also Lax pairs for $\PII$ equations equivalent to those presented in \cite{Adler_Sokolov_2020_1}.

In the scalar case, one can apply to the system \eqref{syscal} the following transformation:
\begin{gather}
    \label{treps}
    \begin{aligned}
    z 
    &= \frac14 \varepsilon^{-3} - \varepsilon \, x, \quad 
    &
    u (z)
    &= - \frac14 \varepsilon^{- 3} - \varepsilon^{-1} \, f (x), \quad 
    &
    v (z)
    &= - 2 \varepsilon \, g (x),
    \end{aligned}
    \\
    \notag
    \begin{aligned}
    c_1
    &= - \frac{1}{16} \varepsilon^{-6},
    &
    c_2
    &= 2 \theta,
    &
    \theta 
    &\in \mathbb{C},
    \end{aligned}
\end{gather}
to bring it to the form
\begin{gather*}
    \left\{
    \begin{array}{lcl}
    f'
    &=& 2 \varepsilon^2 \brackets{2 f g - x f} - f^2 + g - \frac12 x,
    \\[2mm]
    g'
    &=& 2 \varepsilon^{2}
    \brackets{ - g^2 + x g} + 2 f g + \theta
    .
    \end{array}
    \right.
\end{gather*}
Passing to the limit $\varepsilon \to 0$, we obtain the system
\begin{equation} \label{eq:P2sys}
    \left\{
    \begin{array}{lcl}
         f'
         &=& - f^2 + g - \frac12 x, 
         \\[2mm]
         g'
         &=& 2 f g + \theta,
    \end{array}
    \right.
\end{equation}
which is equivalent to the \PPainleve-2 equation 
\begin{equation}
    \label{eq:P2}
    y''
    = 2 y^3 + x y + \brackets{\theta - \frac12}
\end{equation}
for $y (x) = f (x)$.

This degeneration procedure can be extend to the Lax pair \eqref{eq:scalP4_Laxpair}. Conjugating \eqref{eq:scalP4_Laxpair} by the matrix $\tilde{g}~=~\small \begin{pmatrix} 1 & - u \\ 0 & 1 \end{pmatrix}$, we get
\begin{align*}
    B (z, \zeta)
    &= 
    \begin{pmatrix}
    1 & 2 u 
    \\[0.9mm] 
    0 & -1
    \end{pmatrix}
    \zeta 
    + 
    \begin{pmatrix}
    - u + \frac12 v - z & 0 
    \\[0.9mm]
    1 & u - \frac12 v + z
    \end{pmatrix}
    ,
    \\[2mm]
    A (z, \zeta)
    &= 
    \begin{pmatrix}
    -1 & - 2 u 
    \\[0.9mm] 
    0 & 1
    \end{pmatrix}
    \zeta 
    + 
    \begin{pmatrix}
    u + z & - u v + u^2 + 2 z u - c_1 
    \\[0.9mm]
    -1 & - u - z
    \end{pmatrix}
    + \frac12 
    \begin{pmatrix}
    \frac12 c_2 & 0 
    \\[0.9mm]
    v & - \frac12 c_2
    \end{pmatrix}
    \zeta^{-1}
    .
\end{align*}
Let us make
the transformation \eqref{treps}, supplemented by 
\begin{equation} \label{zzet}
\zeta = 2 \varepsilon \lambda,
\end{equation}
in the relation \eqref{eq:isomLax}.  This leads to the $\varepsilon$-dependent pair 
\begin{align*}
    B (x, \lambda)
    &= 
    \begin{pmatrix}
    - 2 \varepsilon^2 
    & 
    \varepsilon^{-1} + 4 \varepsilon f
    \\[0.9mm]
    0 
    & 
    2 \varepsilon^2
    \end{pmatrix}
    \lambda 
    + 
    \begin{pmatrix}
    - f + \varepsilon^2 g - \varepsilon^2 x
    & 
    0 
    \\[0.9mm]
    - \varepsilon
    & 
    f - \varepsilon^2 g + \varepsilon^2 x
    \end{pmatrix}
    ,
    \\[2mm]
    A (x, \lambda)
    &= 
    \begin{pmatrix}
    - 4 \varepsilon^2 
    & 
    2 \brackets{
    \varepsilon^{-1} 
    + 4 \varepsilon f
    }
    \\[0.9mm]
    0 
    & 
    4 \varepsilon^2
    \end{pmatrix}
    \lambda 
    \\[1mm]
    & \qquad
    + 
    \begin{pmatrix}
    - 2 f - 2 \varepsilon^2 x
    & 
    - 4 \varepsilon f  \brackets{
    g - x
    }
    + \varepsilon^{-1} \brackets{
    2 f^2 - g + x
    }
    \\[0.9mm]
    - 2 \varepsilon
    & 
    2 f + 2 \varepsilon^2 x
    \end{pmatrix}
    + 
    \begin{pmatrix}
    \frac12 \theta & 0 
    \\[0.9mm]
    - \varepsilon g & - \frac12 \theta
    \end{pmatrix}
    \lambda^{-1}
    ,
\end{align*}
in which, after conjugation by the matrix $ g = \diag (1, \varepsilon^{- 1}) $, one can pass to the limit $ \varepsilon \to 0 $ to obtain the pair 
\begin{gather} \label{eq:P2_Laxpair}
    \begin{aligned}
    B (x, \lambda)
    &= 
    \begin{pmatrix}
    0
    & 
    1
    \\[0.9mm]
    0 
    & 
    0
    \end{pmatrix}
    \lambda 
    + 
    \begin{pmatrix}
    - f
    & 
    0 
    \\[0.9mm]
    - 1
    & 
    f
    \end{pmatrix}
    ,
    \\[2mm]
    A (x, \lambda)
    &= 
    \begin{pmatrix}
    0
    & 
    2
    \\[0.9mm]
    0 
    & 
    0
    \end{pmatrix}
    \lambda 
    + 
    \begin{pmatrix}
    - 2 f
    & 
    2 f^2 - g + x
    \\[0.9mm]
    - 2
    & 
    2 f
    \end{pmatrix}
    + 
    \begin{pmatrix}
    \frac12 \theta & 0 
    \\[0.9mm]
    - g & - \frac12 \theta
    \end{pmatrix}
    \lambda^{-1}
    \end{aligned}
\end{gather}
for the $\text{P}_2$ system \eqref{eq:P2sys}. Here we replace $z$ by $x$ and $\zeta$ by $\lambda$ in the formula \eqref{eq:isomLax}. The pair \eqref{eq:P2_Laxpair} is just the Harnad–Tracy–Widom pair \cite{harnad1993hamiltonian} up to a scaling.

We are going to describe similar relations between matrix systems \ref{eq:case1tail_can} -- \ref{eq:case5tail_can} and $\PII$ equations  \ref{P20} -- \ref{P22}.
All these equations have the form
\begin{align}
    y'' 
    &= \kappa [y, y']
    + 2 y^3 + x y + b_1 y + y b_2 + a,
    &
    \kappa
    &\in \mathbb{C}, \,\,
    a, b_1, b_2
    \in \Mat_n (\mathbb{C}).
\end{align}
Such a matrix equation can be rewritten as the system of two equations
\begin{align} \label{eq:P2matsys}
     &\left\{
    \begin{array}{lcl}
         f'
         &=& - f^2 + g - \frac12 x \, \mathbb{I} - c_1,  
         \\[2mm]
         g'
         &=& 2 g f + \beta [g, f] 
         + c_2 f + f c_3 + c_4,
    \end{array}
    \right.
    &
    \beta
    &\in \mathbb{C}, \,\,
    c_i 
    \in \Mat_n (\mathbb{C}).
\end{align}
Here \,\, $f(x) = y (x)$,\,\, $\kappa = - 1 - \beta$, \,\, $b_1 = c_2 + (2 + \beta) c_1$, \,\, $b_2 = c_3 - \beta c_1$, and $a = c_4 - \frac12 \, \mathbb{I}$.

\subsection{Case \texorpdfstring{$\alpha = \beta = -1$}{alpha = beta = -1}}
Let us apply transformations \eqref{treps}, \eqref{zzet} taking together with
\begin{align*}
    h_1
    &= 4 \varepsilon b,
    &
    \gamma_1 
    &= - \frac{1}{16} \varepsilon^{-6},
    &
    \gamma_2 
    &= 2 \theta,
    &
    b 
    &\in \Mat_n (\mathbb{C}),
    &
    \theta 
    &\in \mathbb{C}
\end{align*}
to the pair \eqref{eq:case1_Laxpair}. Then, after the passage to the limit  $\varepsilon \to 0$, we obtain matrices
\begin{gather} \label{eq:P20_HTWpair}
\begin{aligned}
    B (x, \lambda)
    &= 
    \begin{pmatrix}
    0
    & 
    \mathbb{I}
    \\[0.9mm]
    0 
    & 
    0
    \end{pmatrix}
    \lambda 
    + 
    \begin{pmatrix}
    - f
    & 
    0 
    \\[0.9mm]
    - \mathbb{I}
    & 
    f
    \end{pmatrix}
    ,
    \\[2mm]
    A (x, \lambda)
    &= 
    \begin{pmatrix}
    0
    & 
    2 \, \mathbb{I}
    \\[0.9mm]
    0 
    & 
    0
    \end{pmatrix}
    \lambda 
    + 
    \begin{pmatrix}
    - 2 f
    & 
    2 f^2 - g + x \, \mathbb{I} + 2 b
    \\[0.9mm]
    - 2 \, \mathbb{I}
    & 
    2 f
    \end{pmatrix}
    + 
    \begin{pmatrix}
    \frac12 \theta \, \mathbb{I} & 0 
    \\[0.9mm]
    - g & - \frac12 \theta \, \mathbb{I}
    \end{pmatrix}
    \lambda^{-1}
    ,
\end{aligned}
\end{gather}
which define the Lax pair for the $\text{P}_2^0$ system
\begin{equation} \label{eq:P20sys}
    \left\{
    \begin{array}{lcl}
         f'
         &=& - f^2 + g - \frac12 x \, \mathbb{I} - b, 
         \\[2mm]
         g'
         &=& f g + g f + \theta \, \mathbb{I}
    \end{array}
    \right.
\end{equation}
of the form \eqref{eq:P2matsys}. 

\subsection{Case \texorpdfstring{$\alpha = 0$, $\beta = -2$}{alpha = 0, beta = -2}}
As in the previous case, for the limiting transition in the pair \eqref{eq:case3_Laxpair}, we supplement the change of variables \eqref{treps}, \eqref{zzet} with the following transformations:
\begin{align*}
    h_2
    &= 2 a - \frac{1}{16} \varepsilon^{-6} \, \mathbb{I},
    &
    \gamma_1
    &= \frac{1}{16} \varepsilon^{-6},
    &
    a 
    &\in \Mat_n (\mathbb{C}),
\end{align*}
and perform the shift of the form \eqref{shift} with $\kappa_3 = \frac{1}{64} \varepsilon^{-6} \, \mathbb{I}$. Then in the limit $\varepsilon \to 0 $ we get the pair  
\begin{align*}
    B (x, \lambda)
    &= 
    \begin{pmatrix}
    0
    & 
    \mathbb{I}
    \\[0.9mm]
    0 
    & 
    0
    \end{pmatrix}
    \lambda 
    + 
    \begin{pmatrix}
    - f
    & 
    0 
    \\
    - \mathbb{I}
    & 
    f
    \end{pmatrix}
    ,
    \\[2mm]
    A (x, \lambda)
    &= 
    \begin{pmatrix}
    0
    & 
    2 \, \mathbb{I}
    \\[0.9mm]
    0 
    & 
    0
    \end{pmatrix}
    \lambda 
    + 
    \begin{pmatrix}
    - 2 f
    & 
    2 f^2 - g + x \, \mathbb{I}
    \\[0.9mm]
    - 2 \, \mathbb{I}
    & 
    2 f
    \end{pmatrix}
    + 
    \begin{pmatrix}
    0 & 0 
    \\[0.9mm]
    - g & - f g + g f - a
    \end{pmatrix}
    \lambda^{-1}
    ,
\end{align*}
for the $\text{P}_2^1$ system 
\begin{equation*}
    \left\{
    \begin{array}{lcl}
         f'
         &=& - f^2 + g - \frac12 x \, \mathbb{I}, 
         \\[2mm]
         g'
         &=& 2 f g + a
    \end{array}
    \right.
\end{equation*}
of the form \eqref{eq:P2matsys}. This system is equivalent to the \ref{P21} equation for $y (x) = f (x)$.

\begin{remark} The system is also equivalent to the equation
\begin{gather} \label{eq:P34case4}
    w''
    = \frac12 (w' - a) w^{-1} (w' + a)
    + 2 w^2 - x w,
\end{gather}
for the variable $w(x) = g(x)$. In the scalar case it coincides with the \text{\rm$\text{P}_{34}$} equation
\begin{align} \label{eq:P34scalarcase}
    &&
    w''
    &= \frac12 \brackets{(w')^2 - a^2} w^{-1} + 2 w^2 - x w,
    &
    a 
    &\in \mathbb{C}.
    &&
\end{align}
\end{remark}

\subsection{Case \texorpdfstring{$\alpha = 0$, $\beta = -3$}{alpha = 0, beta = -3}}
Consider the transformations \eqref{treps}, \eqref{zzet} padded with formulas
\begin{align*}
    h_2 
    &= a - \frac18 \varepsilon^{-2} b - \frac12 \gamma_1 \, \mathbb{I},
    &
    h_1 
    &= - \frac83 \varepsilon b,
    &
    \gamma_1
    &= \frac{1}{8} \varepsilon^{-6},
    &
    \kappa_6 
    &= \frac{1}{4} \varepsilon^{-2} \, \mathbb{I},
    &
    a, b 
    &\in \Mat_n (\mathbb{C}),
\end{align*}
and with the shift $g \mapsto g + \frac23 b$.  
Then the pair \eqref{eq:case5_Laxpair} in the limit $\varepsilon \to 0$ takes the form 
\begin{align*}
    B (x, \lambda)
    &= 
    \begin{pmatrix}
    0
    & 
    \mathbb{I}
    \\[0.9mm]
    0 
    & 
    0
    \end{pmatrix}
    \lambda 
    + 
    \begin{pmatrix}
    0
    & 
    - \frac13 b
    \\[0.9mm]
    - \mathbb{I}
    & 
    2 f
    \end{pmatrix}
    ,
    \\[2mm]
    A (x, \lambda)
    &= 
    \begin{pmatrix}
    0
    & 
    2 \, \mathbb{I}
    \\[0.9mm]
    0 
    & 
    0
    \end{pmatrix}
    \lambda 
    + 
    \begin{pmatrix}
    - 2 f
    & 
    2 f^2 - g + x \, \mathbb{I} + \frac23 b
    \\[0.9mm]
    - 2 \, \mathbb{I}
    & 
    2 f
    \end{pmatrix}
    \\[1mm]
    & \qquad \qquad
    + 
    \begin{pmatrix}
    \frac23 b f + \frac12 a 
    & 
    \frac13 b g 
    - \frac23 b f^2
    - \frac13 x b 
    - \frac49 b^2
    \\[0.9mm]
    - g + \frac23 b
    & 
    - f g + g f 
    - \frac43 b f + \frac23 f b - \frac12 a
    \end{pmatrix}
    \lambda^{-1}
    .
\end{align*}
The relation \eqref{eq:isomLax}  is equivalent to the system
\begin{align*}
    &\left\{
    \begin{array}{lcl}
         f'
         &=& - f^2 + g - \frac12 x \, \mathbb{I} - b, 
         \\[2mm]
         g'
         &=& 3 f g - g f - 2 [f, b] + a,
    \end{array}
    \right.
    &
    [b, a]
    &= 2 b,
\end{align*}
of the form \eqref{eq:P2matsys} corresponding to the \ref{P22} equation for $y (x) = f (x)$. 

Thus, we found the Lax pairs for matrix \Painleve equations \ref{P20} -- \ref{P22}. Our pairs are gauge equivalent to those presented in \cite{Adler_Sokolov_2020_1}. In particular, the Irfan pair \cite{Irfan_2012, Adler_Sokolov_2020_1}
\begin{gather*}
    \begin{aligned}
    B (x, \mu)
    &= 
    \begin{pmatrix}
    \mu \, \mathbb{I} & - f 
    \\[0.9mm] 
    - f & - \mu \, \mathbb{I}
    \end{pmatrix},
    \\[2mm]
    A (x, \mu)
    &= - 4 B(x, \mu) \, \mu 
    +
    \begin{pmatrix}
    2 f^2 + x \, \mathbb{I} + 2 b
    &
    - 2 f^2 + 2 g - x \, \mathbb{I} - 2 b
    \\[0.9mm]
    2 f^2 - 2 g + x \, \mathbb{I} + 2 b
    &
    - 2 f^2 - x \, \mathbb{I} - 2 b
    \end{pmatrix}
    + \brackets{\theta - \frac12}
    \begin{pmatrix}
    0 & \mathbb{I} \\ \mathbb{I} & 0
    \end{pmatrix}
    \mu^{-1} 
    \end{aligned}
\end{gather*}
for system $\text{P}_2^0$ \eqref{eq:P20sys} 
can be reduced to \eqref{eq:P20_HTWpair} by conjugation with the matrix
$g~=~\small\begin{pmatrix} \mu^{-\frac12} \, \mathbb{I} & - \mu^{\frac12} \, \mathbb{I} \\ \mu^{-\frac12} \, \mathbb{I} & \phantom{-} \mu^{\frac12} \, \mathbb{I} \end{pmatrix}$ 
(the so-called Fabri transform, see \cite{joshi2009linearization}), where $\lambda = - \mu^2$.
 
Using the Fabri transformations it is not difficult to convert our pairs to pairs of the Flashka-Newell type \cite{flaschka1980monodromy}. However, they look more cumbersome.

\medskip

\section{Conclusion}
\subsection{\texorpdfstring{$\text{P}_4$}{P4}--equation with non-commutative independent variable}

The systems \ref{eq:case1tail_can} -- \ref{eq:case5tail_can}
as well as their Lax pairs found in Section \ref{sect4} are polynomial and therefore
make sense if we replace the matrix variables with elements of an  associative unital algebra  ${\cal A}$ over $\mathbb{C}$. In the case of systems \ref{eq:case1tail_can}, \ref{eq:case3tail_can} we can take for this algebra the trivial central extension  of the free associative algebra with generators $u$, $v$, and $h_i$ by the element $z$. For system \ref{eq:case5tail_can}
the algebra is generated by $z$, $u$, $v$, $h_1$,  $h_2$ and by the relation $[h_2, h_1] = - 2 h_1$.

In this approach a system \eqref{3rootTail} is replaced by the corresponding derivation $D$ of ${\cal A}$, which acts on the generators~as
$$
D(b_i)=D(c_i)=0, \quad D(u)= - u^2 + 2 \,u v  + \alpha (u v - v u) - 2 z u + b_1 u+u b_2+b_3 v+v b_4+b_5, 
$$
$$ D(v)=-  v^2 + 2 \, v u + \beta (v u - u v) + 2 z v+c_1 v+v c_2 +c_3 u +u c_4+c_5, \qquad D(z)=1.
$$
Formula \eqref{eq:isomLax} means that 
$
    D(A)=B_\zeta+[B,A].
$
It would be interesting to find additional commutator relations that can be imposed on the  generators of ${\cal A}$ in the case of systems \ref{eq:case1tail_can}~--~\ref{eq:case5tail_can} \, \cite{nagoya2008quantum, mikhailov2020quantisation,inamasu2021matrix}.

For systems of Painlev\'e type with non-commutative independent variable (cf. \cite{Retakh_Rubtsov_2010}) the algebra  ${\cal A}$ is the free associative algebra with generators $\bar z$, $u$, and $v$. Since in such models we have $D(\bar z)=1$, in the matrix representation the non-commutative variable $\bar z$ may be replaced by $z \, \mathbb{I} + b$, where $z$ is the commutative independent variable and $b$ is an arbitrary constant matrix.  Therefore, any system with the non-commutative $\bar z$ that has a Lax representation \eqref{eq:isomLax} generates the corresponding system with commutative $z$. The converse is not always true. However, if in the pair \eqref{eq:case1_Laxpair} we add the term $(-\zeta-z) \, \mathbb{I}$ to $B$ and  $(\zeta-z) \, \mathbb{I}$ to $A$ by a shift \eqref{shift}, the result can be written as 
\begin{gather*}
    \begin{aligned}
    B (\bar{z}, \zeta)
    &= 
    \begin{pmatrix}
    0 & 0 \\[0.9mm] 0 & - 2
    \end{pmatrix}
    \zeta 
    + 
    \begin{pmatrix}
    v - 2 \bar{z} & u v + \gamma_1
    \\[0.9mm]
    1 & 0
    \end{pmatrix},
    \\[2mm]
    A (\bar{z}, \zeta)
    &= 
    \begin{pmatrix}
    0 & 0 \\[0.9mm] 0 & 2
    \end{pmatrix}
    \zeta
    + 
    \begin{pmatrix}
    0 & - u v - \gamma_1
    \\[0.9mm]
    -1 & - 2 \bar{z}
    \end{pmatrix}
    + \frac12 
    \begin{pmatrix}
    u v + \frac12 \gamma_2 & - u v u - \gamma_2 u
    \\[0.9mm]
    v & - v u - \frac12 \gamma_2
    \end{pmatrix}
    \zeta^{-1}
    ,
    \end{aligned}
\end{gather*}
where $\bar z = z \, \mathbb{I} - \frac{1}{2} h_1$. This is the Lax pair for the \ref{eq:case1tail_can} system 
\begin{align*}
    &&
    &\left\{
    \begin{array}{lcl}
         u' 
         &=& - u^2 + u v + v u - 2 \bar z u
         + \gamma_1,
         \\[2mm]
         v' 
         &=& - v^2 + v u + u v + 2 v \bar z
         + \gamma_2,
    \end{array}
    \right.
    &
    \gamma_1, \gamma_2 
    &\in \mathbb{C},
    &&
\end{align*}
with non-commutative independent variable $\bar{z}$.

For slightly more general system \cite{Bobrova_Sokolov_2020_1}
\begin{equation} \label{roubtsov}
    \left\{
    \begin{array}{lcl} 
    u'
    &=& - u^2 + u v + v u + (k - 2) \, \bar zu - k\, u \bar z + \gamma_1,
    \\[2mm]
    v'
    &=& - v^2 + v u + u v + k\, \bar z v - (k - 2)\, v \bar z + \gamma_2,
    \end{array}
    \right.
\end{equation}
where $k\in  \mathbb{C},$ the Lax pair  has the form 
\begin{gather*}
    \begin{aligned}
    B (\bar{z}, \zeta)
    &= 
    \begin{pmatrix}
    0 & 0 \\[0.9mm] 0 & - 2
    \end{pmatrix}
    \zeta 
    + 
    \begin{pmatrix}
    v + (k - 2) \bar{z} & u v + \gamma_1
    \\[0.9mm]
    1 & k \bar{z}
    \end{pmatrix},
    \\[2mm]
    A (\bar{z}, \zeta)
    &= 
    \begin{pmatrix}
    0 & 0 \\[0.9mm] 0 & 2
    \end{pmatrix}
    \zeta
    + 
    \begin{pmatrix}
    0 & - u v - \gamma_1
    \\[0.9mm]
    -1 & - 2 \bar{z}
    \end{pmatrix}
    + \frac12 
    \begin{pmatrix}
    u v + \frac12 \gamma_2 & - u v u - \gamma_2 u
    \\[0.9mm]
    v & - v u - \frac12 \gamma_2
    \end{pmatrix}
    \zeta^{-1}
    .
    \end{aligned}
\end{gather*}

\subsection{Outlook} 

Possibly, our approach can be applied to the classification of matrix polynomial systems of  $\text{P}_6$--type. It is natural to expect that the limiting transitions (see Section \ref{limit}) relate them to other matrix Painlev\'e type systems. Besides, one can try to generalize recent results related to  matrix $\text{P}_1$ and $\text{P}_2$ equations \cite{gordoa2010backlund, Pic, inamasu2021matrix, cafasso2014non, manas2021matrix}. In these papers 
hierarchies of higher matrix Painlev\'e equations, non-commutative monodromy
surfaces (for a matrix $\text{P}_2$-equation there exist five non-commutative monodromy surfaces 
\cite{Bertola2018}), auto-B\"acklund transformations,  and non-commutative orthogonal polynomials are investigated. 

\subsubsection*{Acknowledgements}

The authors are grateful to V. Adler, V. Poberezhny and V. Roubtsov for useful discussions. The research of the second author was carried out under the State Assignment 0029-2021-0004 (Quantum field theory) of the Ministry of Science and Higher Education of the Russian Federation. The first author was partially supported by the International Laboratory of Cluster Geometry NRU HSE, RF Government grant № 075-15-2021-608.
  
\bibliographystyle{plain}
\bibliography{bib}

\begin{thebibliography}{10}

\bibitem{Adler_Sokolov_2020_1}
V.~E. Adler and V.~V. Sokolov.
\newblock On matrix {P}ainlev{\'e} {II} equations.
\newblock {\em Theoret. and Math. Phys.},
  207(2):\href{https://doi.org/10.4213/tmf10027}{188--201}, 2021.
\newblock \href{https://arxiv.org/abs/2012.05639}{arXiv:2012.05639}.

\bibitem{Balandin_Sokolov_1998}
S.~P. Balandin and V.~V. Sokolov.
\newblock On the {P}ainlev{\'e} test for non-{A}belian equations.
\newblock {\em Physics letters A},
  246(3-4):\href{https://www.sciencedirect.com/science/article/abs/pii/S0375960198003363?via\%3Dihub}{267--272},
  1998.

\bibitem{Bertola2018}
M.~Bertola, M.~Cafasso, and V.~Rubtsov.
\newblock Noncommutative {P}ainlev{\'e} equations and systems of {C}alogero
  type.
\newblock {\em Communications in Mathematical Physics},
  363(2):\href{https://link.springer.com/article/10.1007\%2Fs00220--018--3210--0}{503--530},
  2018.
\newblock \href{https://arxiv.org/abs/1710.00736}{arXiv:1710.00736}.

\bibitem{Bobrova_Sokolov_2020_1}
I.~A. Bobrova and V.~V. Sokolov.
\newblock On matrix {P}ainlev{\'e}-4 equations. {P}art 1:
  {P}ainlev{\'e}-{K}ovalevskaya test.
\newblock {\em arXiv preprint
  \href{https://arxiv.org/abs/2107.11680}{arXiv:2107.11680}}, 2021.

\bibitem{manas2021matrix}
A.~Branquinho, A.~Moreno, and M.~Mañas.
\newblock Matrix biorthogonal polynomials: {E}igenvalue problems and
  non-{A}belian discrete {P}ainlev\'e equations: a {R}iemann–{H}ilbert
  problem perspective.
\newblock {\em Journal of Mathematical Analysis and Applications},
  494(2):\href{https://doi.org/10.1016/j.jmaa.2020.124605}{124605}, 2021.

\bibitem{cafasso2014non}
M.~Cafasso and D.~Manuel.
\newblock Non-commutative {P}ainlev{\'e} equations and {H}ermite-type matrix
  orthogonal polynomials.
\newblock {\em Communications in Mathematical Physics},
  326(2):\href{https://doi.org/10.1007/s00220--013--1853--4}{559--583}, 2014.

\bibitem{flaschka1980monodromy}
H.~Flaschka and A.~C. Newell.
\newblock Monodromy- and spectrum-preserving deformations {I}.
\newblock {\em Communications in Mathematical Physics},
  76(1):\href{https://link.springer.com/article/10.1007/BF01197110}{65--116},
  1980.

\bibitem{gambier1910equations}
B.~Gambier.
\newblock Sur les {\'e}quations diff{\'e}rentielles du second ordre et du
  premier degr{\'e} dont l'int{\'e}grale g{\'e}n{\'e}rale est {\`a} points
  critiques fixes.
\newblock {\em Acta Mathematica},
  33(1):\href{https://projecteuclid.org/journals/acta--mathematica/volume--33/issue--none/Sur--les--\%c3\%a9quations--diff\%c3\%a9rentielles--du--second--ordre--et--du--premier/10.1007/BF02393211.full}{1--55},
  1910.

\bibitem{gordoa2010backlund}
P.~R. Gordoa, A.~Pickering, and Z.~N. Zhu.
\newblock B\"acklund transformations for a matrix second {P}ainlev\'e equation.
\newblock {\em Physics Letters A},
  374(34):\href{https://doi.org/10.1016/j.physleta.2010.06.034}{3422--3424},
  2010.

\bibitem{Pic}
P.~R. Gordoa, A.~Pickering, and Z.~N. Zhu.
\newblock On matrix {P}ainlev{\'e} hierarchies.
\newblock {\em Journal of Differential Equations},
  261(2):\href{https://www.sciencedirect.com/science/article/pii/S0022039616300092}{1128--1175},
  2016.

\bibitem{harnad1993hamiltonian}
J.~Harnad, C.~A. Tracy, and H.~Widom.
\newblock Hamiltonian structure of equations appearing in random matrices.
\newblock In {\em Low-Dimensional Topology and Quantum Field Theory}, pages
  \href{https://link.springer.com/chapter/10.1007/978--1--4899--1612--9_21}{231--245}.
  Springer, 1993.
\newblock \href{https://arxiv.org/abs/hep-th/9301051}{arXiv:9301051}.

\bibitem{inamasu2021matrix}
K.~Inamasu and H.~Kimura.
\newblock Matrix hypergeometric functions, semi-classical orthogonal
  polynomials and quantum {P}ainlev{\'e} equations.
\newblock {\em Integral Transforms and Special Functions},
  32(5-8):\href{https://doi.org/10.1080/10652469.2020.1833878}{528--544}, 2021.

\bibitem{Irfan_2012}
M.~Irfan.
\newblock Lax pair representation and {D}arboux transformation of
  noncommutative {P}ainlev{\'e}’s second equation.
\newblock {\em Journal of Geometry and Physics},
  62(7):\href{https://doi.org/10.1016/j.geomphys.2012.01.008}{1575--1582},
  2012.
\newblock \href{https://arxiv.org/abs/1201.0900}{arXiv:1201.0900}.

\bibitem{Jimbo_Miwa_1981}
M.~Jimbo and T.~Miwa.
\newblock Monodromy preserving deformation of linear ordinary differential
  equations with rational coefficients. {II}.
\newblock {\em Physica D: Nonlinear Phenomena},
  2(3):\href{https://www.sciencedirect.com/science/article/abs/pii/016727898190021X}{407--448},
  1981.

\bibitem{joshi2009linearization}
N.~Joshi, A.~V. Kitaev, and P.~A. Treharne.
\newblock On the linearization of the first and second {P}ainlev{\'e}
  equations.
\newblock {\em Journal of Physics A: Mathematical and Theoretical},
  42(5):\href{https://doi.org/10.1088/1751--8113/42/5/055208}{055208}, 2009.
\newblock \href{https://arxiv.org/abs/0806.0271v1}{arXiv:0806.0271v1}.

\bibitem{joshi2007linearization}
N.~Joshi, A.V. Kitaev, and P.A. Treharne.
\newblock On the linearization of the {P}ainlev{\'e} {III--VI} equations and
  reductions of the three-wave resonant system.
\newblock {\em Journal of Mathematical Physics},
  48(10):\href{https://aip.scitation.org/doi/10.1063/1.2794560}{103512}, 2007.
\newblock \href{https://arxiv.org/abs/0706.1750v3}{arXiv:0706.1750v3}.

\bibitem{mikhailov2020quantisation}
A.~V. Mikhailov.
\newblock Quantisation ideals of nonabelian integrable systems.
\newblock {\em Russian Mathematical Surveys},
  75(5):\href{https://iopscience.iop.org/article/10.1070/RM9966}{978–980},
  2020.
\newblock \href{https://arxiv.org/abs/2009.01838}{arXiv:2009.01838}.

\bibitem{nagoya2008quantum}
H.~Nagoya, B.~Grammaticos, A.~Ramani, et~al.
\newblock Quantum {P}ainlev{\'e} equations: from {C}ontinuous to discrete.
\newblock {\em SIGMA. Symmetry, Integrability and Geometry: Methods and
  Applications}, 4:\href{https://www.emis.de/journals/SIGMA/2008/051/}{051},
  2008.

\bibitem{Retakh_Rubtsov_2010}
V.~S. Retakh and V.~V. Rubtsov.
\newblock Noncommutative {T}oda {C}hains, {H}ankel {Q}uasideterminants and
  {P}ainlev{\'e} {II} {E}quation.
\newblock {\em Journal of Physics. A, Mathematical and Theoretical},
  43(50):\href{https://iopscience.iop.org/article/10.1088/1751--8113/43/50/505204}{505204},
  2010.
\newblock \href{https://arxiv.org/abs/1007.4168}{arXiv:1007.4168}.

\end{thebibliography}

\end{document}